\documentclass[apj]{emulateapj}
\usepackage[colorlinks=true,citecolor=blue,urlcolor=cyan]{hyperref}
\usepackage{apjfonts,ulem,color}
\usepackage{amssymb, amsmath}
\newcommand\degree{\degr}
\newcommand\degrees\degree

\DeclareSymbolFont{UPM}{U}{eur}{m}{n}
\DeclareMathSymbol{\umu}{0}{UPM}{"16}
\let\oldumu=\umu
\renewcommand\umu{\ifmmode\oldumu\else\math{\oldumu}\fi}
\newcommand\micro{\umu}
\renewcommand\micron{\micro m}
\newcommand\microns \micron

\renewcommand\arcsec[0]{$^{\prime\prime}$}

\let\oldsim=\sim
\renewcommand\sim{\ifmmode\oldsim\else\math{\oldsim}\fi}
\let\oldpm=\pm
\renewcommand\pm{\ifmmode\oldpm\else\math{\oldpm}\fi}
\newcommand\by{\ifmmode\times\else\math{\times}\fi}

\newbox{\wdbox}
\renewcommand\c{\setbox\wdbox=\hbox{,}\hspace{\wd\wdbox}}
\renewcommand\i{\setbox\wdbox=\hbox{i}\hspace{\wd\wdbox}}

\newcount\timect
\newcount\hourct
\newcount\minct
\newcommand\now{\timect=\time \divide\timect by 60
         \hourct=\timect \multiply\hourct by 60
         \minct=\time \advance\minct by -\hourct
         \number\timect:\ifnum \minct < 10 0\fi\number\minct}

\catcode`@=11

\newcommand\comment[1]{}
\newcommand\commenton{\catcode`\%=14}
\newcommand\commentoff{\catcode`\%=12}

\renewcommand\math[1]{$#1$}
\newcommand\mathshifton{\catcode`\$=3}
\newcommand\mathshiftoff{\catcode`\$=12}

\comment{the backslash is necessary}

\comment{alignment tab}

\let\atab=&
\newcommand\atabon{\catcode`\&=4}
\newcommand\ataboff{\catcode`\&=12}

\let\oldmsp=\sp
\let\oldmsb=\sb
\def\sp#1{\ifmmode
           \oldmsp{#1}%
         \else\strut\raise.85ex\hbox{\scriptsize #1}\fi}
\def\sb#1{\ifmmode
           \oldmsb{#1}%
         \else\strut\raise-.54ex\hbox{\scriptsize #1}\fi}
\newbox\@sp
\newbox\@sb
\def\sbp#1#2{\ifmmode%
           \oldmsb{#1}\oldmsp{#2}%
         \else
           \setbox\@sb=\hbox{\sb{#1}}%
           \setbox\@sp=\hbox{\sp{#2}}%
           \rlap{\copy\@sb}\copy\@sp
           \ifdim \wd\@sb >\wd\@sp
             \hskip -\wd\@sp \hskip \wd\@sb
           \fi
        \fi}
\def\msp#1{\ifmmode
           \oldmsp{#1}
         \else \math{\oldmsp{#1}}\fi}
\def\msb#1{\ifmmode
           \oldmsb{#1}
         \else \math{\oldmsb{#1}}\fi}
\def\supon{\catcode`\^=7}
\def\supoff{\catcode`\^=12}
\def\subon{\catcode`\_=8}
\def\suboff{\catcode`\_=12}
\def\supsubon{\supon \subon}
\def\supsuboff{\supoff \suboff}

\newcommand\actcharon{\catcode`\~=13}
\newcommand\actcharoff{\catcode`\~=12}

\newcommand\paramon{\catcode`\#=6}
\newcommand\paramoff{\catcode`\#=12}

\comment{And now to turn us totally on and off...}

\newcommand\reservedcharson{\commenton \mathshifton \atabon \supsubon \actcharon
	\paramon}

\newcommand\reservedcharsoff{\commentoff \mathshiftoff \ataboff
	\supsuboff \actcharoff \paramoff}

\catcode`@=12
\reservedcharsoff

\reservedcharson

\comment{ Must have ONLY ONE of these... trust these macros, they work

}

\newenvironment{packed_enum}{
\begin{enumerate}
   \setlength{\itemsep}{1pt}
   \setlength{\parskip}{0pt}
   \setlength{\parsep}{0pt}
}{\end{enumerate}}	

\newcommand{\squishlist}{
 \begin{list}{$\bullet$}
  { \setlength{\itemsep}{1pt}
     \setlength{\parsep}{0pt}
     \setlength{\topsep}{3pt}
     \setlength{\partopsep}{0pt}
     \setlength{\leftmargin}{2.0em}
     \setlength{\labelwidth}{1.5em}
     \setlength{\labelsep}{0.5em} } }

\newcommand{\squishend}{
  \end{list}  }

\reservedcharson
\actcharon

\shorttitle{Transiting Exoplanet Studies and Community Targets for {\em JWST}'s ERS Program}
\shortauthors{Stevenson {\em et al.}}

\submitted{}

\begin{document}

\title{Transiting Exoplanet Studies and Community Targets for\\ {\em JWST}'s Early Release Science Program}

\author{Kevin B.\ Stevenson\altaffilmark{1,41}}
\author{Nikole K.\ Lewis\altaffilmark{2}}
\author{Jacob L.\ Bean\altaffilmark{1}}
\author{Charles Beichman\altaffilmark{3}}
\author{Jonathan Fraine\altaffilmark{4}}
\author{Brian M.\ Kilpatrick\altaffilmark{5}}
\author{J.E.\ Krick\altaffilmark{6}}
\author{Joshua D.\ Lothringer\altaffilmark{7}}
\author{Avi M.\ Mandell\altaffilmark{8}}
\author{Jeff A.\ Valenti\altaffilmark{2}}
\author{Eric Agol\altaffilmark{9}}
\author{Daniel Angerhausen\altaffilmark{10,42}}
\author{Joanna K.\ Barstow\altaffilmark{11}}
\author{Stephan M.\ Birkmann\altaffilmark{12}}
\author{Adam Burrows\altaffilmark{13}}
\author{David Charbonneau\altaffilmark{14}}
\author{Nicolas B.\ Cowan\altaffilmark{15}}
\author{Nicolas Crouzet\altaffilmark{16}}
\author{Patricio E.\ Cubillos\altaffilmark{17}}
\author{S.M.\ Curry\altaffilmark{18}}
\author{Paul A.\ Dalba\altaffilmark{19}}
\author{Julien de Wit\altaffilmark{20}}
\author{Drake Deming\altaffilmark{21}}
\author{Jean-Michel D\'esert\altaffilmark{22}}
\author{Ren\'e Doyon\altaffilmark{23}}
\author{Diana Dragomir\altaffilmark{1}}
\author{David Ehrenreich\altaffilmark{24}}
\author{Jonathan J.\ Fortney\altaffilmark{25}}
\author{Antonio Garc{\'i}a Mu{\~n}oz\altaffilmark{26}}
\author{Neale P.\ Gibson\altaffilmark{27}}
\author{John E.\ Gizis\altaffilmark{28}}
\author{Thomas P.\ Greene\altaffilmark{29}}
\author{Joseph Harrington\altaffilmark{30}}
\author{Kevin Heng\altaffilmark{31}}
\author{Tiffany Kataria\altaffilmark{32}}
\author{Eliza M.-R.\ Kempton\altaffilmark{33}}
\author{Heather Knutson\altaffilmark{34}}
\author{Laura Kreidberg\altaffilmark{1}}
\author{David Lafreni\`ere\altaffilmark{23}}
\author{Pierre-Olivier Lagage\altaffilmark{35}}
\author{Michael R.\ Line\altaffilmark{29}}
\author{Mercedes Lopez-Morales\altaffilmark{14}}
\author{Nikku Madhusudhan\altaffilmark{36}}
\author{Caroline V.\ Morley\altaffilmark{25}}
\author{Marco Rocchetto\altaffilmark{37}}
\author{Everett Schlawin\altaffilmark{4}}
\author{Evgenya L.\ Shkolnik\altaffilmark{38}}
\author{Avi Shporer\altaffilmark{39,41}}
\author{David K.\ Sing\altaffilmark{32}}
\author{Kamen O.\ Todorov\altaffilmark{40}}
\author{Gregory S.\ Tucker\altaffilmark{5}}
\author{Hannah R.\ Wakeford\altaffilmark{10,42}}

\affil{\sp{1} Department of Astronomy and Astrophysics, University of Chicago, 5640 S Ellis Ave, Chicago, IL 60637, USA}
\affil{\sp{2} Space Telescope Science Institute, 3700 San Martin Drive, Baltimore, MD 21218, USA}
\affil{\sp{3} NASA Exoplanet Science Institute, California Institute of Technology, Jet Propulsion Laboratory, Pasadena, CA, USA}
\affil{\sp{4} Steward Observatory, University of Arizona, Tucson, AZ 85721, USA}
\affil{\sp{5} Department of Physics, Brown University, Providence, RI 02912, USA}
\affil{\sp{6} Spitzer Science Center, California Institute of Technology, Pasadena, CA 91106, USA}
\affil{\sp{7} Lunar \& Planetary Laboratory, University of Arizona, Tucson, AZ 85721, USA}
\affil{\sp{8} Solar System Exploration Division, NASA Goddard Space Flight Center, Greenbelt, MD 20771, USA}
\affil{\sp{9} University of Washington, Box 351580, Seattle, WA 98195, USA}
\affil{\sp{10} NASA Goddard Space Flight Center, Greenbelt, MD 20771, USA}
\affil{\sp{11} Department of Physics, University of Oxford, Denys Wilkinson Building, Keble Road, Oxford, OX1 3RH, UK}
\affil{\sp{12} European Space Agency / Space Telescope Science Institute, 3700 San Martin Drive, Baltimore MD, 21218, USA}
\affil{\sp{13} Department of Astrophysical Sciences, Peyton Hall, Princeton University, Princeton, NJ 08544, USA}
\affil{\sp{14} Harvard-Smithsonian Center for Astrophysics, 60 Garden Street, Cambridge, MA 02138, USA}
\affil{\sp{15} McGill Space Institute, 3550 rue University, Montr{\'e}al, QC H3A 2A7, Canada}
\affil{\sp{16} Dunlap Institute for Astronomy \& Astrophysics, University of Toronto, Toronto, Ontario, Canada}
\affil{\sp{17} Space Research Institute, Austrian Academy of Sciences, Schmiedlstrasse 6, A-8042 Graz, Austria}
\affil{\sp{18} Space Sciences Laboratory, 7 Gauss Way, University of California, Berkeley, Berkeley, CA 94720, USA}
\affil{\sp{19} Department of Astronomy, Boston University, Boston, MA 02215, USA}
\affil{\sp{20} Department of Earth, Atmospheric and Planetary Sciences, MIT, Cambridge, MA 02139, USA}
\affil{\sp{21} Department of Astronomy, University of Maryland, College Park, MD 20742, USA}
\affil{\sp{22} Anton Pannekoek Institute for Astronomy, University of Amsterdam, The Netherlands}
\affil{\sp{23} Institut de Recherche sur les Exoplan\`etes, D\'epartement de Physique, Universit\'e de Montr\'eal, C.P. 6128, Succ. Centre-vile, Montr\'eal, QC H3C 3J7, Canada}
\affil{\sp{24} Observatoire de l'Universit{\'e} de Gen{\`e}ve, 51 chemin des Maillettes, 1290 Versoix, Switzerland}
\affil{\sp{25} Department of Astronomy and Astrophysics, University of California, Santa Cruz, CA 95064, USA}
\affil{\sp{26} Zentrum f{\"u}r Astronomie und Astrophysik, Technische Universit{\"a}t Berlin, D-10623 Berlin, Germany}
\affil{\sp{27} Astrophysics Research Centre, School of Mathematics and Physics, Queens University Belfast, Belfast BT7 1NN, UK}
\affil{\sp{28} Department of Physics and Astronomy, University of Delaware, Newark, DE 19716, USA}
\affil{\sp{29} NASA Ames Research Center, Space Science and Astrobiology Division, Moffett Field, CA 94035, USA}
\affil{\sp{30} Planetary Sciences Group, Department of Physics, University of Central Florida, Orlando, FL 32816-2385, USA}
\affil{\sp{31} University of Bern, Center for Space and Habitability, Sidlerstrasse 5, CH-3012, Bern, Switzerland}
\affil{\sp{32} Astrophysics Group, School of Physics, University of Exeter, Stocker Road, Exeter EX4 4QL, UK}
\affil{\sp{33} Department of Physics, Grinnell College, Noyce Science Building, Grinnell, IA 50112, USA}
\affil{\sp{34} Division of Geological and Planetary Sciences, California Institute of Technology, Pasadena, CA 91125, USA}
\affil{\sp{35} Paris-Saclay University, Irfu/AIM, CEA Saclay, F-91191 Gif-sur-Yvette, France}
\affil{\sp{36} Institute of Astronomy, University of Cambridge, Madingley Road, Cambridge CB3 0HA, UK}
\affil{\sp{37} Department of Physics and Astronomy, University College London, London, NW1 2PS, UK}
\affil{\sp{38} School for Earth and Space Exploration, Arizona State University, 781 S Terrace Rd, Tempe, AZ 85281, USA}
\affil{\sp{39} Jet Propulsion Laboratory, California Institute of Technology, 4800 Oak Grove Drive, Pasadena, CA 91109, USA}
\affil{\sp{40} Institute for Astronomy, ETH Z{\"u}rich, Wolfgang-Pauli-Strasse 27, 8093 Z{\"u}rich, Switzerland}
\affil{\sp{41} NASA Sagan Fellow}
\affil{\sp{42} USRA NASA Postdoctoral Program Fellow}

\email{kbs@uchicago.edu}

\vspace*{3.0\baselineskip}
\begin{abstract}
The {\em James Webb Space Telescope (JWST)} will likely revolutionize transiting exoplanet atmospheric science due to a combination of its capability for continuous, long duration observations and its larger collecting area, spectral coverage, and spectral resolution compared to existing space-based facilities. However, it is unclear precisely how well {\em JWST} will perform and which of its myriad instruments and observing modes will be best suited for transiting exoplanet studies.  In this article, we describe a prefatory {\em JWST} Early Release Science (ERS) Cycle~1 program that focuses on testing specific observing modes to quickly give the community the data and experience it needs to plan more efficient and successful transiting exoplanet characterization programs in later cycles.  We propose a multi-pronged approach wherein one aspect of the program focuses on observing transits of a single target with all of the recommended observing modes to identify and understand potential systematics, compare transmission spectra at overlapping and neighboring wavelength regions, confirm throughputs, and determine overall performances.  In our search for transiting exoplanets that are well suited to achieving these goals, we identify 12 objects (dubbed ``community targets'') that meet our defined criteria.  Currently, the most favorable target is WASP-62b because of its large predicted signal size, relatively bright host star, and location in {\em JWST}'s continuous viewing zone.  Since most of the community targets do not have well-characterized atmospheres, we recommend initiating preparatory observing programs to determine the presence of obscuring clouds/hazes within their atmospheres.  Measurable spectroscopic features are needed to establish the optimal resolution and wavelength regions for exoplanet characterization.  Other initiatives from our proposed ERS program include testing the instrument brightness limits and performing phase-curve observations.  The latter are a unique challenge compared to transit observations because of their significantly longer durations.  Using only a single mode, we propose to observe a full-orbit phase curve of one of the previously characterized, short-orbital-period planets to evaluate the facility-level aspects of long, uninterrupted time-series observations.
\end{abstract}

\section{Introduction}
\label{sec:intro}

\begin{figure*}[tb]
\centering
\includegraphics[width=1.0\linewidth]{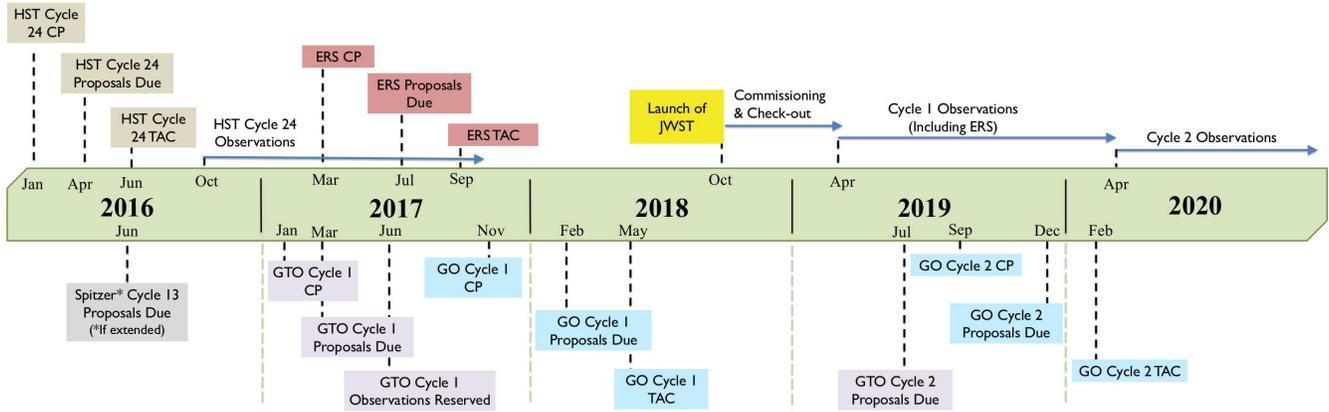}
\caption{\label{fig:timeline}{
Current timeline of preparatory, proposal, and observing events for {\em JWST}.  The one-year proprietary period of GTO/GO Cycle~1 data and the timing of the Cycle~2 proposal deadlines necessitates the establishment of an ERS program.  Given the advancing ERS proposal deadline, preparatory observing programs must be completed as early as possible during {\em HST} Cycle~24 and {\em Spitzer} Cycle~13 (if funding is extended).  See {\href{http://www.stsci.edu/jwst/science/ers}{http://www.stsci.edu/jwst/science/ers}} for the most up-to-date timeline.
}}
\end{figure*}

The {\em James Webb Space Telescope (JWST)} is currently on schedule to launch in October of 2018.  With the commissioning and check-out phases expected to last $\sim$6 months, Cycle~1 programs will commence as early as April of 2019.  This will include Guaranteed Time Observations (GTOs) and General Observer (GO) programs, which will have a standard proprietary period of one year from their observation dates.  With restricted access to key data sets, only a small fraction of the community would initially be able to assess the performance of {\em JWST}'s instruments.  Further slowing scientific progress, the Cycle~2 proposal deadlines are currently scheduled for July and December of 2019 for GTO and GO proposals, respectively (see Figure~\ref{fig:timeline}); therefore, the general community would not be able to write well-informed proposals until Cycle~3 at the earliest.  To solve this problem, the {\em JWST} Advisory Committee\footnote{\href{http://www.stsci.edu/jwst/advisory-committee}{http://www.stsci.edu/jwst/advisory-committee}} recommended the creation of an Early Release Science (ERS) program.

The purpose of the ERS program is to provide open access to a broad suite of {\em JWST} science observations as early as possible in Cycle~1\footnote{\href{http://www.stsci.edu/jwst/science/ers}{http://www.stsci.edu/jwst/science/ers}}.  The program will seed initial discovery, quickly build community experience with {\em JWST}, and inform the preparation of Cycle~2 proposals.
Current plans envisage approved ERS programs that will:
\begin{packed_enum}
\item Be published before the GO Cycle~1 call for proposals,
\item Total $\sim$500 hours of telescope time,
\item Be reviewed and selected by peer-review and executed by the community,
\item Span key {\em JWST} observing modes, data analysis challenges, and science areas,
\item Be science driven to provide the building blocks for future programs,
\item Have no proprietary period to encourage fast turn-around and analyses from multiple teams,
\item Be among the first Cycle~1 observations, and
\item Include the rapid delivery of science-enabling products to the community by ERS teams.
\end{packed_enum}
For this final point, examples of science-enabling products could range from simple observing cookbooks with descriptions of the various instruments systematics to full data reduction pipelines that produce time-series spectra.

Significant effort has gone into simulating {\em JWST} spectra, typically including known and anticipated sources of random and systematic error, for the purpose of predicting the telescope's on-sky performance and scientific output \citep[e.g., ][]{Deming2009, Batalha2015, Barstow2015, Greene2016}.  The implementation of one or more transiting exoplanet ERS programs will provide a means to quickly evaluate these predictions and revise expectations where necessary.  In order for the community to optimize signal-to-noise estimates and telescope time requests to most efficiently achieve their scientific goals in future observing cycles, it is important to have an ERS program that promptly measures the performances of all of the recommended observing modes, so that we can identify and understand potential systematics and determine which modes are best suited for various science cases at overlapping wavelength regions.

The breakdown of this paper is as follows.  In Section~\ref{sec:jwst}, we describe the {\em JWST} instruments, their recommended exoplanet observing modes, and potential systematics.  Section~\ref{sec:targ} presents the criteria to be labeled a community target and identifies several exoplanets suitable for ERS programs.  In Section~\ref{sec:ers}, we consider several hypothetical exoplanet ERS programs and what they might achieve.  We discuss preparatory programs in Section~\ref{sec:prep} and, finally, summarize our findings in Section~\ref{sec:sum}.

\newpage
\section{{\em JWST} INSTRUMENT MODES AND SYSTEMATICS}
\label{sec:jwst}

\begin{figure}[tb]
\centering
\includegraphics[width=1.0\linewidth]{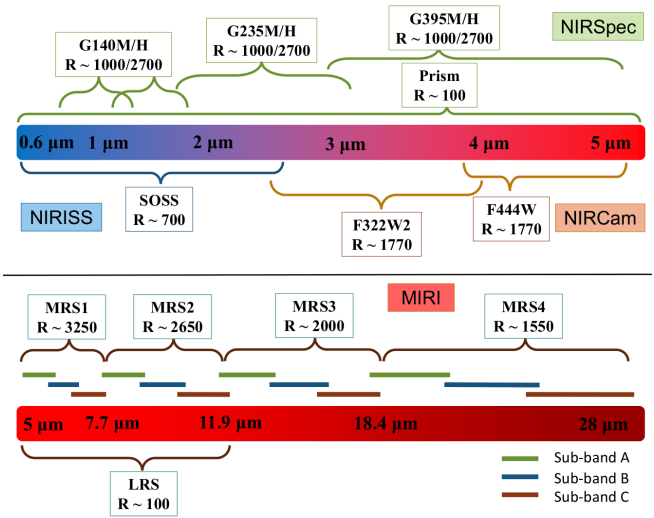}
\caption{\label{fig:modes}{
Recommended spectroscopic observing modes for transiting exoplanets.  For objects brighter than $J~\sim~11$, at least three visits will be required to achieve complete spectral coverage from 0.6 to 5.0 {\microns}.  NIRSpec's low-resolution prism mode may be a viable single-visit option for fainter targets.  MIRI's MRS mode uses a dichroic to simultaneously observe in all four channels, but can only acquire spectral data from one sub-band (A, B, or C) at any given time.  Therefore, the MRS mode requires three visits for complete wavelength coverage.
}}
\end{figure}

Transiting exoplanet studies can utilize all four {\em JWST} scientific instruments.  These include the Near-InfraRed Camera \citep[NIRCam, PI: Marcia Rieke, ][]{Beichman2012}, the Near-InfraRed Spectrograph \citep[NIRSpec, PI: Peter Jakobsen, ][]{Ferruit2012}, the Near Infrared Imager and Slitless Spectrograph \citep[NIRISS, PI: Ren\'e Doyon, ][]{Doyon2012}, and the Mid-Infrared Instrument \citep[MIRI, PIs: George Rieke \& Gillian Wright, ][]{Kendrew2015, Rieke2015, Wells2015}.  Here, we provide a brief description of each instrument as they relate to transiting exoplanets, discuss how they can be utilized in an exoplanet ERS program, and justify why such programs are necessary.  For a more detailed description of the instruments and systematics, see \citet{Beichman2014} and references within.  Figure~\ref{fig:modes} provides a graphical representation of the recommended spectroscopic observing modes at various wavelengths.

\subsection{Instrument Modes For Transiting Exoplanets}

NIRCam consists of two modules that can view the same field at different wavelengths through the use of a dichroic beamsplitter.  From 2.4 to 5.0 {\microns}, a grism can be paired with one of two broadband filters (2.4 -- 4.0 and 3.9 -- 5.0 {\microns}) to perform slitless spectroscopy.  Simultaneously, NIRCam can perform photometry in the range of 0.7 -- 2.4 {\microns} using a narrow, medium, wide, or double wide filter for targets as bright as K$\sim$6 when aided by the use of a defocussing lens.

With NIRSpec, the primary observing modes for transit spectroscopy of exoplanets will utilize the 1.6{\arcsec}${\times}$1.6{\arcsec} aperture over four wavelength regions (0.7 -- 1.2, 1.0 -- 1.8, 1.7 -- 3.1, and 2.9 -- 5.2 {\microns}).  Each region can be observed at medium or high resolution ($R \sim 1000$ and $\sim 2700$, respectively).  For targets fainter than $J~\sim~10$, a low-resolution ($R \sim 100$) prism is available from 0.6 to 5.3 {\microns} without saturation.  Brighter targets can be observed with this mode by saturating (and discarding) certain wavelength regions.

NIRISS has a Single Object Slitless Spectroscopy (SOSS) mode that utilizes a crossed-dispersed grism to obtain simultaneous wavelength coverage from 0.6 to 2.5 {\microns} with $R \sim 700$.  A weak cylindrical lens broadens the spectrum in the spatial direction to a width of 20 -- 25 pixels, thus enabling the observation of targets as bright as J~$\sim$~7.

MIRI provides wavelength coverage from 5.0 to 28 {\microns} in two spectroscopic modes.  The low-resolution spectrograph (LRS, $R \sim 100$) can simultaneously acquire transit data from 5.0 to 12 {\microns} either with or without a slit.  The medium-resolution spectrograph (MRS, $R = 1300 - 3700$) can span the full wavelength range but requires four different integral field units (IFUs, 5.0 -- 7.7, 7.7 -- 11.9, 11.9 -- 18.4, 18.4 -- 28.3 {\microns}) and, because the spectra are interleaved, three visits for complete wavelength coverage.  The MRS mode is likely to have significant data reduction and systematic modeling challenges due to complexities in the instrument design, undersampling of the point spread function, the potential for image slicer and slit losses, and based on previous attempts to obtain and analyze time-series data with IFUs \citep{Angerhausen2006}.  Although MRS is the only spectroscopic option available at wavelengths $>12$ {\microns}, broadband photometry is also feasible.

\subsection{Telescope and Instrument Systematics}

It is vital to make a thorough test of the instruments early in the mission because it is likely that systematics will dominate the signals in the data. {\em JWST} and its instruments constitute a general purpose observatory, and while a lot of work has recently gone in to optimizing and characterizing the instruments for transit observations, the facility was not originally designed for this purpose. Therefore, we expect that systematics like some of those seen in {\em Hubble Space Telescope (HST)} and {\em Spitzer Space Telescope} transit observations could also be seen in {\em JWST} data.  These systematics might arise from properties of the telescope like pointing jitter, drift, and flexure driven by thermal changes; properties of the detectors like inter- and intra-pixel sensitivity variations, persistence or charge trapping, non-linearity, and instability of the readout electronics; properties of the instruments like slit losses, variations of the throughput over the field of view, spectral and spatial pixel sampling, and optical contamination; and operational requirements that interrupt the observations.
We discuss some of these issues below and provide examples from current facilities to further motivate the need for a thorough assessment of {\em JWST}'s capabilities in exoplanet ERS programs.

Early in its life, {\em HST} encountered unexpected vibration issues in its original solar arrays that led to line-of-sight jitter.  Although its arrays were later replaced, the spacecraft still undergoes thermal breathing on a timescale of $\sim 30$~minutes.  The primary source of these systematics is the large thermal gradients experienced when entering and exiting Earth's shadow.  {\em JWST} will not orbit the Earth, so it will not undergo such extreme temperature variations, but large spacecraft motions due to pointing changes may require a significant settling time that may still not fully nullify line-of-sight jitter.  As an example, {\em Spitzer} takes $\sim 30$~minutes to settle subsequent to large changes in pointing, after which source positions undergo jitter (or wobble) with an amplitude of 0.08 pixels and a slow linear drift of 0.01 pixels per hour \citep{Ingalls2016}.  The wobble is caused by a battery heater cycling on and off while the long-term drift is caused by a discrepancy between the spacecraft's instantaneous velocity aberration and the on-board aberration correction that only takes place at the start of each observation.  

{\em JWST} will have a comparable jitter to pixel-scale ratio as that of {\em Spitzer}/IRAC \citep{Beichman2014}, but its impact will vary between instruments.  Those with smaller spectral extent (or height) along their spatial directions will encompass fewer pixels and thus be more susceptible to inter- and intra-pixel sensitivity variations.  The former is due to an imprecise flat field and the latter is thought to be the result of a non-uniform quantum efficiency across the pixel surface.  Intra-pixel sensitivity variations are the primary systematic in {\em Spitzer}'s InSb detector arrays and methods to precisely model this systematic took several years to develop \citep[e.g.,][]{Ballard2010b, Stevenson2011, Lewis2013, Deming2015}.  Recent work by the community \citep{Ingalls2016} has shown that we now have a good understanding of {\em Spitzer} systematics and can achieve consistent results using multiple techniques.

Although NIRCam, NIRSpec, and NIRISS are functionally different instruments, they operate at overlapping near-infrared wavelength regions using similar Teledyne HgCdTe ``Hawaii'' detectors.  Since {\em HST}/WFC3 also uses a HgCdTe Hawaii detector, we can anticipate some of the instrument systematics that we are likely to encounter.   For example, at fluences $>40,000$~e\sp{-} per pixel, WFC3 experiences an increase in intensity (similar to a ramp or hook) after each buffer dump \citep[e.g., ][]{Berta2012, Deming2013, Swain2013, Wilkins2014}.  For similar reasons, but independent of flux, WFC3 also exhibits an {\em HST} orbit-long ramp that is readily apparent in most spatial scan data \citep[e.g., ][]{Kreidberg2014, Wakeford2016}.  Because WFC3 is a slitless spectrograph, it is not subject to slit losses.  However, NIRSpec's 1.6{\arcsec} square aperture may be subject to slit losses in the event of larger-than-expected telescope jitter or drift.

MIRI uses a Si:As array that is a successor to {\em Spitzer}'s IRAC detectors at 5.8 and 8.0 {\microns}.  IRAC's redder channel exhibited a well-studied, time-dependent rising ramp \citep[e.g., ][]{Harrington2007, Knutson2009b, Agol2010, Stevenson2011} while the bluer channel's falling ramp was similar but less-studied.  As a result, we can leverage past experience with {\em Spitzer} to better understand and remove what are likely to be similar systematics in MIRI.

Ultimately, there is no transit-spectroscopy-specific error budget for {\em JWST} or any of its instrument modes. Until we can make on-sky measurements, it is unclear which instrument and observing-mode combination will achieve the best performance with minimal systematics.  Measuring the instruments' relative on-sky performances can most effectively be achieved by observing a single target orbiting a quiet, moderately bright host star in all recommended modes and wavelength regions.  By observing a common source with a quiet host star, the community will be able to directly compare the capabilities of each instrument, investigate potential offsets between exoplanet transmission/emission spectra at overlapping wavelengths, and establish a list of best observing practices from the beginning.  By not pushing the limits of each instrument, we will be able to apply standard detector read patterns (e.g., reset -- read -- read...) and investigate typical instrument behaviors that will apply to the majority of transiting exoplanet observations.


\section{COMMUNITY TARGETS}
\label{sec:targ}

Here we identify exoplanets that are best suited to achieving the goals of the ERS program.  Because ERS data will have no proprietary period and are intended to provide the building blocks for future programs, we refer to these exoplanets as ``community targets.''  Most exoplanets do not qualify as community targets because they do not meet the necessary criteria.  Specifically, we assert that a community target should have the following attributes:
\begin{packed_enum}
\item A high ecliptic latitude ($|\beta|>45${\degrees}),
\item A short orbital period ($P<10$ days),
\item A well-constrained orbital solution and planet mass,
\item A relatively bright and quiet host star ($J<10.5$, $\log R'_{HK}\lesssim -4.8$), and
\item A transmission spectrum with measurable spectroscopic features ($\Delta D>50$ ppm/H).
\end{packed_enum}
A high ecliptic latitude and short orbital period are necessary to ensure a long visibility window ($\gtrsim 6$ months) with a reasonable number of transit opportunities ($\gtrsim 20$).  Unexpected delays can push back the launch date or commencement of Cycle~1 by several months; therefore, a community target cannot have a highly restrictive visibility window.  Also, a planet with a relatively short orbital period is necessary to complete observations in all instrument modes and wavelength regions within a reasonable time frame ($<3$ months).

A well-constrained orbital solution includes a precise ephemeris to minimize transit time uncertainties with {\em JWST} and, desirably, a known eccentricity for future secondary-eclipse observations.  Furthermore, a planet mass constraint is required to interpret a transmission spectrum and draw conclusions about the atmosphere.  Without the presence of transit timing variations (TTVs) to make a mass determination, we must rely on the radial velocity technique, which requires a relatively bright and quiet host star.  The spectroscopic light curve signal-to-noise ratio also benefits from having a bright star, while a quiet host star minimizes confusion between the desired signal and non-Gaussian, time-correlated astrophysical variations.  Finally, it is important that the {\em JWST} instruments are sensitive to spectroscopic features within a transmission spectrum to properly characterize the atmosphere and determine the best resolution.  We estimate the signal size per scale height, $\Delta D \sim \frac{2HR_P}{R_S^2}$ where $H = \frac{k\sb{B}T\sb{eq}}{\mu g}$, assuming a cloud-free atmosphere at constant $T$\sb{eq} and a mean molecular weight of 2.2~u.

\begin{figure}[tb]
\centering
\includegraphics[width=1.0\linewidth]{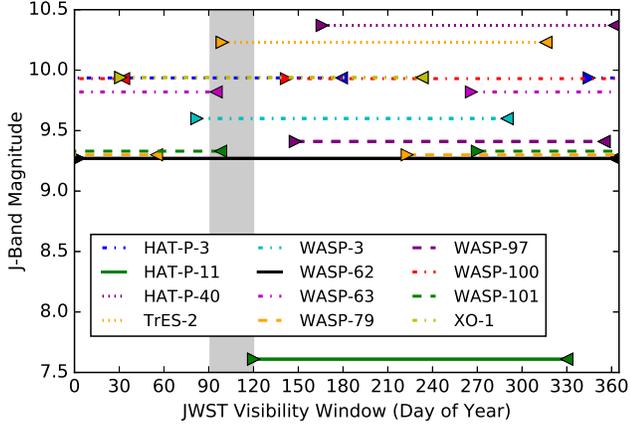}
\caption{\label{fig:viswindow}{
Comparison of J-band magnitude and {\em JWST} visibility window for 12 community targets.  HAT-P-11 (solid green line) is a bright, but active star and WASP-62 (solid black line) is in {\em JWST}'s continuous viewing zone.  In the event that HAT-P-11b and WASP-62b are unsuitable community targets, there are eight additional systems with $J<10$ and at least three community targets that are visible at any given time.  The gray region depicts the nominal start month for Cycle~1.
}}
\end{figure}

Tables~\ref{tab:planets} and~\ref{tab:stars} list the planet and stellar properties, respectively, of 12 systems that meet the criteria for becoming a community target.  If compelling, new targets are discovered in the near future, they could be added to this list.  Figure~\ref{fig:viswindow} compares host star $J$-band magnitudes against their {\em JWST} visibility windows.  HAT-P-11 and WASP-62 are the brightest stars shown; the latter is in {\em JWST}'s continuous viewing zone (CVZ).

\begin{table*}[tb]
\centering
\caption{\label{tab:planets} 
Planet and Observation Properties}
\begin{tabular}{cccccccccc}
    \hline
    \hline      
    Name        & Period    &$T$\sb{eq} & $\log g$  & Tr. Depth & Tr. Dur.  & Signal Size\tablenotemark{a}   
                                                                                            & Vis. Window       & \# Transits   & Score\tablenotemark{b}\\
                & (Days)    & (K)       & (dex)     & (\%)      & (hr)      & (ppm/H)       &                   &               &       \\
    \hline
    HAT-P-3b    & 2.900     & 1160      & 3.26      & 1.23      & 2.1       &  90           & Dec 11 -- Jun 28  & 69            & 3     \\
    HAT-P-11b   & 4.888     &  870      & 3.06      & 0.33      & 2.3       &  60           & Apr 30 -- Nov 26  & 44            & 2     \\
    HAT-P-40b   & 4.457     & 1760      & 2.71      & 0.65      & 6.1       & 140           & Jun 15 -- Dec 31  & 45            & 1     \\
    TrES-2b     & 2.471     & 1500      & 3.30      & 1.57      & 1.8       & 100           & Apr 09 -- Nov 12  & 89            & 3     \\
    WASP-3b     & 1.847     & 1990      & 3.48      & 1.03      & 2.8       &  60           & Mar 24 -- Oct 17  & 112           & 3     \\
    WASP-62b    & 4.412     & 1430      & 2.86      & 1.23      & 3.8       & 180           & Jan 01 -- Dec 31  & 83            & 5     \\
    WASP-63b    & 4.378     & 1530      & 2.66      & 0.61      & 5.3       & 150           & Sep 23 -- Apr 05  & 45            & 4     \\
    WASP-79b    & 3.662     & 1760      & 2.88      & 1.15      & 3.8       & 170           & Aug 11 -- Feb 24  & 54            & 4     \\
    WASP-97b    & 2.073     & 1540      & 3.41      & 1.19      & 2.6       &  70           & May 28 -- Dec 20  & 99            & 3     \\
    WASP-100b   & 2.849     & 2200      & 3.24      & 0.76      & 3.8       &  60           & May 22 -- Jan 31  & 90            & 3     \\
    WASP-101b   & 3.586     & 1550      & 2.79      & 1.26      & 2.7       & 240           & Sep 27 -- Apr 08  & 53            & 5     \\
    XO-1b       & 3.942     & 1210      & 3.19      & 1.76      & 2.9       & 120           & Feb 02 -- Aug 21  & 51            & 4     \\
    \hline
\end{tabular}
\tablenotetext{1}{Predicted signal size per scale height assumes a cloud-free atmosphere at constant $T$\sb{eq} and a mean molecular weight of 2.2~u.}
\tablenotetext{2}{This subjective scoring system ranges from 1 to 5, where higher values are given to targets with more favorable parameters.}
\end{table*}

\begin{table*}[tb]
\centering
\caption{\label{tab:stars} 
Stellar Properties}
\begin{tabular}{ccccccc}
    \hline
    \hline      
    Name        & R.A.          & Decl.         & Ecl. Lat. & V-Band    & J-Band    & $\log R'_{HK}$\tablenotemark{a} \\
                & (HH:MM:SS.ss) & (DD:MM:SS.s)  &($\degree$)& (mag)     & (mag)     &               \\
    \hline
    HAT-P-3     & 13:44:22.58   & +48:01:43.2   & +53.0     & 11.6      &  9.9      & -4.904    \\
    HAT-P-11    & 19:50:50.14   & +48:04:49.1   & +66.6     &  9.6      &  7.6      & -4.567    \\
    HAT-P-40    & 22:22:03.10   & +45:27:26.4   & +50.3     & 11.3      & 10.4      & -5.140    \\
    TrES-2      & 19:07:14.04   & +49:18:59.0   & +70.7     & 11.4      & 10.2      & -4.949    \\
    WASP-3      & 18:34:31.62   & +35:39:41.5   & +58.7     & 10.6      &  9.6      & -4.872    \\
    WASP-62     & 05:48:33.59   & -63:59:18.2   & -87.2     & 10.2      &  9.3      & -4.7      \\
    WASP-63     & 06:17:20.75   & -38:19:23.8   & -61.6     & 11.2      &  9.8      &   --      \\
    WASP-79     & 04:25:29.02   & -30:36:01.5   & -51.3     & 10.0      &  9.3      &   --      \\
    WASP-97     & 01:38:25.04   & -55:46:19.4   & -58.4     & 10.6      &  9.4      &   --      \\
    WASP-100    & 04:35:50.32   & -64:01:37.3   & -80.9     & 10.8      &  9.9      &   --      \\
    WASP-101    & 06:33:24.26   & -23:29:10.2   & -46.6     & 10.3      &  9.3      &   --      \\
    XO-1        & 16:02:11.84   & +28:10:10.4   & +47.6     & 11.2      &  9.9      & -4.958    \\
    \hline
\end{tabular}
\tablenotetext{1}{Values from \citet{Knutson2010}, except for HAT-P-40 and WASP-62 (H. Isaacson and A. Triaud 2016, private communications).}
\end{table*}

\begin{figure}[tb]
\centering
\includegraphics[width=1.0\linewidth]{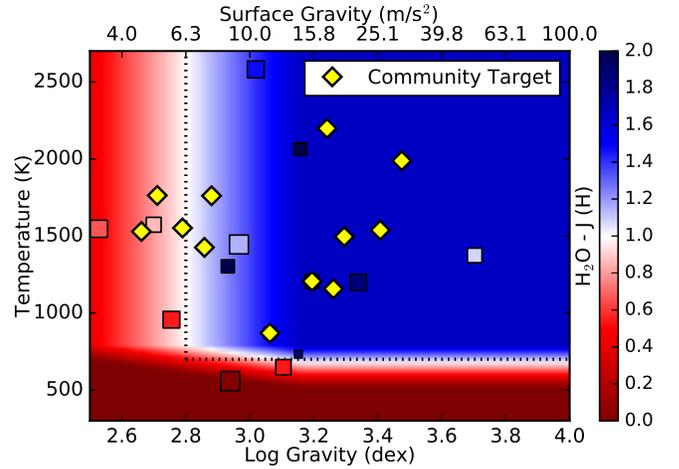}
\caption{\label{fig:teqgravity}{
Expected 1.4 {\micron} water feature strength as functions of equilibrium temperature and surface gravity (background colors).  Blue and red squares represent planets with strong and weak measured water features, respectively, as determined by \citet{Stevenson2016b}.  The dotted lines (where H\sb{2}O~--~J $\sim 1$) delineate the two regimes.  Nine of twelve community targets (yellow diamonds) are expected to have partly cloudy atmospheres or better with measurable spectroscopic features.  HAT-P-40b, WASP-63b, and WASP-101b are the only targets predicted to have muted spectroscopic features due to obscuring clouds.  The model only considers planets with hydrogen-rich atmospheres.
}}
\end{figure}

Figure~\ref{fig:teqgravity} plots the same 12 targets as functions of equilibrium temperature, $T$\sb{eq}, and surface gravity, $g$.  It also displays the measured 1.4 {\micron} water feature strength of 14 planets (two of which overlap) from {\em HST}/WFC3 observations.  Based on the model by \citet{Stevenson2016b}, which suggests that exoplanets with $T$\sb{eq}$>700$~K and $\log g > 2.8$~dex are less likely to have obscuring clouds, nine of the community targets should have measurable spectroscopic features in transmission with {\em HST} and {\em JWST}.  Three exceptions (HAT-P-40b, WASP-63b, and WASP-101b) are more likely to have muted spectroscopic features, though not necessarily flat spectra.

In the subsections below, we discuss each system in detail, including their advantages and disadvantages as community targets in a hypothetical ERS program.  We also rate the targets subjectively on a scale from 1 to 5, where higher values are given to targets with more favorable parameters.  New information can always change the scores of individual targets.  One important aspect to consider when planning exoplanet observations is the transit duration.  Planets with shorter transit durations make for more efficient ERS programs because they require fewer hours for a fixed number of observations.  No community target will require more than one visit per observing mode to measure a signal because preparatory programs (see Section~\ref{sec:prep}) will vet targets with flat transmission spectra.  Other aspects to consider include the timing of the visibility window and the brightness of the host star.  It is important to note that all comparisons below are relative to the other targets on this list.

\subsection{HAT-P-3b}

HAT-P-3b \citep{Torres2007} has the second shortest transit duration, a moderate predicted signal size, and orbits a $J=9.9$ star that is visible from December through June.  \citet{Todorov2013} report on secondary-eclipse measurements of HAT-P-3b using {\em Spitzer} and find elevated levels of red noise in the light curves that may be due to slightly enhanced chromospheric activity from the host star.

{\em Score:} 3

\subsection{HAT-P-11b}

\citet{Fraine2014} report the detection of water vapor in the transmission spectrum of HAT-P-11b \citep{Bakos2010}.  This suggests this Neptune-size object has a predominantly cloud-free atmosphere at the pressure levels probed by {\em HST}/WFC3.  The planet also has a relatively short transit duration (2.3 hr) and a visibility window that begins within {\em JWST}'s nominal Cycle~1 start month.  HAT-P-11 is the brightest host star on the list and would saturate the detector in some observing modes.  The star is also relatively active, causing visible star spot crossing events in the transit light curves \citep{Deming2011, Sanchis-Ojeda2011, Fraine2014}.

{\em Score:} 2

\subsection{HAT-P-40b}

HAT-P-40b \citep{Hartman2012} has the longest transit duration across the faintest host star on the list.  It also has relatively few transits during its visibility window, which does not start until mid-June.  Although the planet's theoretical spectroscopic feature size is 140 ppm/H, clouds may obscure part of that signal \citep{Stevenson2016b}.

{\em Score:} 1

\subsection{TrES-2b}

Like HAT-P-11b, TrES-2b \citep{ODonovan2006} has a short transit duration (1.8 hr, due to a near-grazing transit) and a visibility window that coincides with {\em JWST}'s nominal Cycle~1 start date.  In contrast to HAT-P-11b, TrES-2b's shorter orbital period and deeper transit depth results in twice as many transit opportunities and a $\sim60\%$ larger predicted signal (assuming cloud-free atmospheres).  TrEs-2 is 2.6 mag fainter than HAT-P-11 in the $J$-band and second faintest overall.  {\em HST}/WFC3 observed a single transit of TrES-2b in the less-efficient stare mode; therefore, the reported depths \citep{Ranjan2014} have insufficient precision to definitively detect water vapor or constrain the presence of clouds.  Future observations with spatial scan mode should achieve the necessary precision.  \citet{Kipping2011} and \citet{Raetz2014} present constraints from high-precision {\em Kepler} photometry that reveal to measurable TTVs.

{\em Score:} 3

\subsection{WASP-3b}

WASP-3b \citep{Pollacco2008} has a reasonable transit duration and, due to its short 1.8 day orbital period, has the highest number of transit opportunities.  WASP-3 is the second-brightest host star (after WASP-62) whose visibility window coincides with {\em JWST}'s nominal Cycle~1 start date.  Therefore, should WASP-62b not be a viable target then WASP-3b may present a good option.  Unfortunately, the planet's large mass (2.0 $M$\sb{J}) does result in a small predicted signal size that is comparable to that of HAT-P-11b.  Although there are four transit observations of WASP-3b (three with NICMOS and one stare-mode with WFC3/G102, GO Program 11495), none of their results have been published.  Regardless, we estimate that these data sets have insufficient precision to constrain the presence of clouds.
{\em Spitzer} thermal emission observations at 3.6, 4.5, and 8.0 {\microns} \citep{Rostron2014} reveal moderate levels of correlated noise in one of the residual light curves.

{\em Score:} 3

\subsection{WASP-62b}

WASP-62 \citep{Hellier2012} is the second-brightest host star (after HAT-P-11) and has the distinction of being the only system in {\em JWST}'s CVZ.  This results in a large number of transit opportunities at any time of year.  WASP-62b has the second largest predicted signal size (180 ppm/H) for a cloud-free atmosphere; however, it also has a relatively long transit duration (3.8 hr), thus potentially requiring a sizable number of hours to complete an ERS program.

{\em Score:} 5

\subsection{WASP-63b}

WASP-63b's \citep{Hellier2012} is an inflated planet with a large predicted signal size, but its visibility window opens $\sim6$ months after the nominal Cycle~1 start date.  This could make WASP-63b a good target should {\em JWST} experience any significant delays.  The planet's host star is moderately bright ($J=9.8$), but its 5.3 hr transit duration is the second-longest of all the targets.

{\em Score:} 3

\subsection{WASP-79b}

Similar to WASP-63b, WASP-79b \citep{Smalley2012} has a large predicted signal size and an ill-timed visibility window.  Otherwise, the planet's transit duration and host star brightness are more favorable than those of WASP-63b.  Unpublished {\em Spitzer} eclipse observations at 3.6 and 4.5 {\microns} may reveal clues about WASP-79's activity level.

{\em Score:} 4

\subsection{WASP-97b}

WASP-97 \citep{Hellier2014} is nearly as bright as two other favorable target stars (WASP-62 and WASP-79) and the planet's transit duration is 1.2 hr shorter, but WASP-97b's cloud-free predicted signal size (70 ppm/H) is 2.5$\times$ smaller than these other planets.

{\em Score:} 3

\subsection{WASP-100b}

WASP-100b \citep{Hellier2014} has a large number of transit opportunities due to its long visibility window ($>8$ months) that opens shortly after the nominal Cycle~1 start date.  However, it also has a relatively long transit duration and one of the smallest predicted signal sizes (60 ppm/H), similar to HAT-P-11b and WASP-3b.

{\em Score:} 3

\subsection{WASP-101b}

This highly inflated planet \citep{Hellier2014} has the largest predicted signal size, its 2.7 hr transit duration is quite reasonable, and its host star is relatively bright.  Interestingly, WASP-101b's only weakness may turn out to be an asset.  The planet's visibility window opens $\sim6$ months after the nominal Cycle~1 start date; however, should {\em JWST} experience any significant delays and WASP-62b be cloudy, then WASP-101b may become the most viable community target for an ERS program.

{\em Score:} 5

\subsection{XO-1b}

With reasonable transit duration (2.9~hr), predicted signal size (120~ppm/H), visibility window, and host star brightness ($J=9.9$), none of XO-1b's \citep{McCullough2006} physical characteristics enable it to stand out from the crowd.  It is actually the lack of any downside that makes this system so appealing.  Furthermore, \citet{Deming2013} detect a $\sim 200$~ppm water absorption feature in transmission using {\em HST}/WFC3, thus confirming the absence of obscuring clouds.

{\em Score:} 4

\section{EXOPLANET ERS PROGRAMS}
\label{sec:ers}

There are a multitude of hypothetical exoplanet ERS programs whose data could reveal unexpected instrument systematics, propel new data analysis challenges, and ultimately enhance our understanding of {\em JWST} observing modes.
Below we describe several potential programs and the issues that they might address.
We emphasize that {\em JWST}'s commissioning and check-out phases are not meant for programs such as these and that a total of $\sim$500~hr will be available for ERS programs.

\subsection{Program 1: Simply the Best}

The goal of this program is to identify the best overall {\em JWST} observing modes for exoplanet characterization.  Important factors to consider include the instrument duty cycle (the fraction of time gathering photons) and efficiency (number of visits needed to cover a certain wavelength region), the quality of the spectroscopic light curves (considering both precision and presence of time-correlated noise), and the impact of any instrument systematics (including size and repeatability).  For example, a large but repeatable instrument systematic that is wavelength independent can more readily be modeled and, thus, may result in more precise corrected spectroscopic light curves than those from a similar mode with smaller systematics that exhibit less-predictable behavior. 

The best way to compare {\em JWST} observing modes, while still performing interesting science, is to acquire data contiguously from the same source because the individual observing modes cannot all acquire data simultaneously.  Thus, observing the same exoplanet transiting a quiet star in all recommended modes minimizes potential effects due to variations in source brightness and stability, and simplifies the comparison process.  It also minimizes the prospect of planet variability \citep{Demory2016}.  Observing primary transits (instead of secondary eclipses) ensures that we will measure a signal at all wavelengths and allows us to verify the adequacy of our limb-darkening models, which is important given the unrivaled precision expected with {\em JWST}.

There are good reasons to test as many of the recommended observing modes as possible for each instrument.  First, the amplitude of any systematics may be wavelength dependent and second, with overlapping wavelength regions, we will be able to compare transmission spectra on both absolute and relative scales.  The latter will help identify wavelength sub-regions that do not behave as expected.  For example, we'll be able to precisely identify where the edges of each filter no longer provide reliable results under the assumptions of our analyses.  This, in turn, will help guide the stitching process between spectra from neighboring wavelength regions by minimizing potential conflicts.

NIRCam, NIRSpec, and NIRISS utilize a total of at least seven observing modes (see Figure~\ref{fig:modes}) that are suitable for exoplanet observations.  Together, they obtain complete, overlapping spectra from 0.6 to 5.0 {\microns}.  Thus, with at least seven visits, any single community target can likely fulfill the needs of this program.  For example, WASP-62b would require $\sim 58$~hr of {\em JWST} time to complete Program 1.  This includes two hours of out-of-transit baseline both before and after transit to determine the baseline flux and adequately constrain any instrument systematics, plus 30 minutes per visit for telescope settling (8.3 hr total per visit).  Similar strategies are applied to time-series {\em Spitzer} observations.

\subsection{Program 2: MIRI, MIRI on the Wall}

This program is similar in design to the one above; however, it focuses on the only instrument capable of acquiring data at wavelengths $> 5.0$ {\microns} where important absorption by cloud species may be present.  MIRI's LRS mode and the first two channels of its MRS mode acquire overlapping spectra from 5.0 to 12 {\microns}.  This program can determine if a single visit at low resolution (LRS mode) is sufficient for exoplanet characterization or if three visits at medium resolution (MRS mode) are necessary for complete wavelength coverage.  

Additionally, Program 2 can check for consistency with neighboring NIRCam and NIRSpec spectra at 5.0 {\microns} and MRS Channels 3 and 4 at $> 12$ {\microns}.  To facilitate the comparison at lower wavelengths while maximizing ERS efficiency, this program should use the same target as Program~1 and acquire only transit data.  Due to the synergy between observing modes and programs, an alternate grouping could include LRS as part of Program~1 and the addition of broadband photmetry using MIRI's imager to Program~2.

Four visits are needed to obtain spectra in all of the recommended observing modes while obtaining complete wavelength coverage.  Correspondingly for WASP-62b, Program 2 would require 33 hr.  Any emission spectroscopy or direct imaging measurements would require additional visits.

\subsection{Program 3: The Sky is Not the Limit}

The purpose of this program is to test the brightness limits of the instruments using the fastest read patterns available while still meeting the science requirement of the ERS program.  Program 3 can search for flux-dependent systematic effects and deviations from linearity that might bias exoplanet characterization.  As an example of the former, {\em HST}'s WFC3 detector exhibits a ramp- or hook-like systematic at fluences $>40,000$~e\sp{-} \citep{Berta2012, Wilkins2014}, and for the latter, the {\em Spitzer}/IRAC linearity correction is good to $\sim1\%$, which is insufficient for high-precision measurements such as these.  This program can also validate the fastest read patterns that will eventually be used to characterize the atmospheres of the nearest transiting exoplanets.

Targets would be selected to closely match the reported brightness limits of each detector.  A brighter host star such as HAT-P-11 is likely a reasonable target for NIRISS ($J>8.1$ in its standard mode).  Conversely, a fainter host star (e.g., HAT-P-40, TrES-2, or XO-1) may be suitable for NIRSpec using the low-resolution prism.  Both targets can efficiently test their respective detectors during a single visit by covering large ranges in wavelength and flux (due to sensitivity variations in the response curves).  For these two targets, Program 3 would require $\sim 13$~hr of {\em JWST} time.  NIRCam, which does not offer a broad wavelength option for spectroscopy, has an L-band limit of 3.7 and MIRI's K-band limit using LRS mode is in the range of 3 -- 5 (see Table 4 from \citet{Beichman2014}).  Due to the limited availability of known transiting-planet host stars at these magnitudes, a regular GO or engineering program is better suited to test the limits of these instruments.

\subsection{Program 4: Phase Curves in HD}

This program would seek to observe a single target over its entire orbit, thereby measuring the planet's emission as a function of orbital phase.  The goal is not to compare different observing modes, but rather validate one of the recommended modes for this challenging observation.  A full exoplanet phase curve with {\em JWST} would reveal a multitude of important information, including any day-night variation in composition or thermal structure, the heat redistribution efficiency, and the presence of inhomogeneous cloud cover.  Such a long-duration observation could also reveal systematics and stability issues only apparent over these longer baselines.  

\begin{figure}[tb]
\centering
\includegraphics[width=1.0\linewidth]{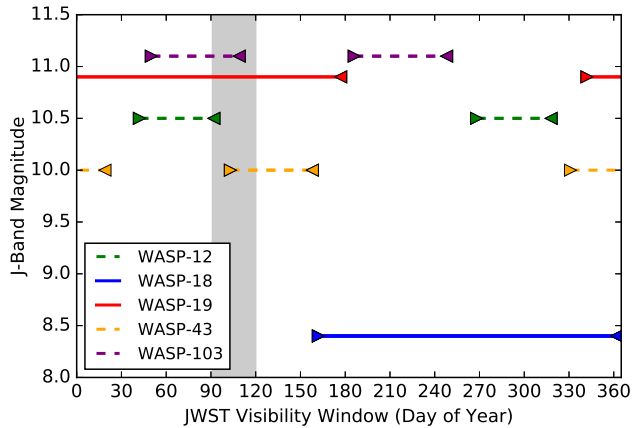}
\caption{\label{fig:viswindow-pc}{
Comparison of J-band magnitude and {\em JWST} visibility window for five potential phase-curve targets.  Together, WASP-18 and WASP-19 (solid lines) offer continuous viewing of at least one target.  One of WASP-43's visibility windows also coincides with {\em JWST}'s nominal start month.  The gray region depicts the nominal start month for Cycle~1.
}}
\end{figure}

Interpretation of the first phase-curve observation would be best served by a target with a short ($\lesssim 24$~hr) orbital period.  This excludes all of our community targets, but a search of short-orbital-period exoplanets with existing {\em HST} or {\em Spitzer} phase curves reveals five promising options: WASP-12b, WASP-18b, WASP-19b, WASP-43b, and WASP-103b (see Figure~\ref{fig:viswindow-pc}).  Standard practices include beginning an observation shortly before secondary eclipse and finishing shortly after the subsequent eclipse to more-reliably anchor the phase curve against any long-term drift in flux.  Therefore, this program will require $\lesssim 30$~hr of continuous telescope time to complete one observation.  

The most interesting observing mode might be NIRSpec's G395M/H grating (2.9 -- 5.2 {\microns}) because it overlaps with {\em warm Spitzer}'s two broadband photometric channels, thus providing an excellent comparison to previously published phase-curve results from a well-studied instrument.  Alternatively, any mode that encompasses 1.1 -- 1.7 {\microns} could prove fruitful for targets whose phase curves have been observed by {\em HST}/WFC3.  The successful completion of Programs 1 and 2 would help guide future phase-curve observations in other modes.

\subsection{Total Time Commitment}

Assuming Programs 1 and 2 select WASP-62b as their community target, the sum of Programs 1 -- 4 is 134 hr, which is $\sim 27\%$ of the total 500 hr allocation for ERS programs.  This percentage is in line with {\em JWST}'s four principal science themes: first light and reionization, the assembly of galaxies, the birth of stars and protoplanetary systems, and planets and origins of life.

\section{PREPARATORY PROGRAMS}
\label{sec:prep}

In preparation for the ERS program with {\em JWST}, we need {\em HST} and {\em Spitzer} observations of the most promising community targets.  The primary goal of these observations should be to identify community targets with predominantly cloud-free atmospheres.  Obscuring clouds/hazes significantly reduce the size of spectroscopic features and limit the amount of information that can be obtained from the observations (such as the optimal resolution and wavelength regions for abundance retrieval studies).  Currently, the presence of clouds/hazes can be determined by measuring the strength of the 1.4 {\micron} water vapor feature  using {\em HST}/WFC3 alone \citep[e.g., ][]{Deming2013, Mandell2013, Kreidberg2014b, Kreidberg2015, Stevenson2016b} or calculating the slope in the transmission spectrum between the {\em Spitzer} and {\em HST}/WFC3 wavelength regions \citep[e.g., ][]{Sing2016}.  Additional goals should include measuring the photometric stability of the host stars, improving ephemerides, searching for nearby companion stars, and providing a baseline for comparison between current facilities and {\em JWST} instruments.  Once one or more suitable community targets have been identified, we recommend follow-up observations using {\em HST}'s COS and STIS instruments at wavelengths not accessible with {\em JWST}.

The ERS proposal submission deadline is expected to be in the summer of 2017; therefore, to allow adequate time for data analyses, publications, and ERS proposal writing, all preparatory observations should be completed by the end of 2016.  Furthermore, since ERS program observations cannot overlap with GTO program observations, we must coordinate with the GTO team leads to avoid duplication issues and ensure that the combination of GTO and ERS programs efficiently covers the diverse array of recommended {\em JWST} observing modes.  The GTO proposal deadline is currently set for April 2017.  To meet these pressing requirements, we recommend initiating {\em HST} and {\em Spitzer} observing programs immediately.

\section{SUMMARY}
\label{sec:sum}

The {\em JWST} Advisory Committee expressed concerns that the dearth of publicly available data prior to the Cycle~2 proposal deadlines would lead to intellectually costly delays both in our understanding of {\em JWST} and its scientific output.  As a solution, they recommended the creation of the Early Release Science program to provide open access to a broad suite of {\em JWST} science observations as early as possible in Cycle~1.

As part of an open discussion at the Enabling Transiting Exoplanet Science with {\em JWST} workshop\footnote{\href{http://www.stsci.edu/jwst/science/exoplanets}{http://www.stsci.edu/jwst/science/exoplanets}} (held 2015 November 16--18 at STScI), the proposition to designate ``community targets'' for one or more exoplanet ERS programs took hold.  Here we have identified 12 potential community targets, the most favorable being WASP-62b because of its large predicted signal size, relatively bright host star, and location in {\em JWST}'s CVZ.  

We have described several key exoplanet ERS programs designed to compare the diverse array of recommended {\em JWST} observing modes, quickly enhance our observing and data analysis experience, and seed initial scientific discovery.  Program~1 advocates for acquiring transits of a single target using seven observing modes over three instruments (NIRCam, NIRSpec, and NIRISS) to identify the best modes for exoplanet characterization at wavelengths $<5.0$~{\microns}.  Program~2 recommends performing a similar assessment with the same target except at longer wavelengths using MIRI's various observing modes.  Program~3 explores testing the brightness limits of some instrument modes using fast read patterns and Program~4 proposes a single phase-curve observation of a target with previously published results.

To determine the optimal resolution and wavelength regions for abundance retrieval studies and, ultimately, meet the science requirements set forth by the ERS program, a community target must have an atmosphere with measurable spectroscopic features (i.e., a non-flat transmission spectrum).  The prevalence of clouds/hazes in exoplanet atmospheres makes identifying viable targets prior to the first GTO proposal deadline a priority.  Therefore, we recommend initiating preparatory observing programs with {\em HST} and {\em Spitzer} to characterize the atmospheres of the most promising community targets by the end of 2016.

\acknowledgments

KBS recognizes support from the Sagan Fellowship Program, supported by NASA and administered by the NASA Exoplanet Science Institute (NExScI).

\clearpage
\bibliography{ms}

\end{document}